\documentclass[english]{article}

\usepackage[colorlinks,bookmarksopen,bookmarksnumbered,allcolors=red]{hyperref}
\usepackage{xcolor}
\usepackage{graphicx}
\usepackage{geometry}
\usepackage{amsmath, amssymb, bm}
\usepackage{authblk}

\definecolor{bluegreen}{RGB}{46,141,131}
\definecolor{darkgreen}{RGB}{46,139,87}
\definecolor{darkred}{RGB}{219,7,61}
\definecolor{darkblue}{RGB}{0,0,137}

\title{\textbf{Functional principal component analysis as an alternative to mixed-effect models for describing sparse repeated measures in presence of missing data}}

\author[,1]{Corentin S\'{e}galas\thanks{Corresponding author: corentin.segalas@u-bordeaux.fr}}
\author[2]{Catherine Helmer}
\author[$\dagger$,1]{Robin Genuer}
\author[,2]{C\'{e}cile Proust-Lima\thanks{Robin Genuer and Cécile Proust-Lima contributed equally as co-last authors}}
\affil[1]{Univ. Bordeaux, INSERM, INRIA, BPH, U1219, F-33000 Bordeaux, France}
\affil[2]{Univ. Bordeaux, INSERM, BPH, U1219, F-33000 Bordeaux, France}

\date{}

\begin{document}

\maketitle

\begin{abstract}
Analyzing longitudinal data in health studies is challenging due to sparse and error-prone measurements, strong within-individual correlation, missing data and various trajectory shapes. While mixed-effect models (MM) effectively address these challenges, they remain parametric models and may incur computational costs. In contrast, Functional Principal Component Analysis (FPCA) is a non-parametric approach developed for regular and dense functional data that flexibly describes temporal trajectories at a potentially lower computational cost. This paper presents an empirical simulation study evaluating the behaviour of FPCA with sparse and error-prone repeated measures and its robustness under different missing data schemes in comparison with MM. The results show that FPCA is well-suited in the presence of missing at random data caused by dropout, except in scenarios involving most frequent and systematic dropout. Like MM, FPCA fails under missing not at random mechanism. The FPCA was applied to describe the trajectories of four cognitive functions before clinical dementia and contrast them with those of matched controls in a case-control study nested in a population-based aging cohort. The average cognitive declines of future dementia cases showed a sudden divergence from those of their matched controls with a sharp acceleration 5 to 2.5 years prior to diagnosis.\newline\newline
\end{abstract}

\section{Introduction}
\label{sec::intro}

In medical research, it is very common to collect in a cohort of participants the repeated measures of various markers at different visits in order to study the trajectories of the underlying biological processes. For example, in HIV research studies, the temporal evolution of CD4 is used to assess the immune system response \cite{boscardin_longitudinal_1998}. In prostate cancer, the prostate specific antigen is a marker used to monitor prostate cancer progression \cite{ilic_prostate_2018}. In cognitive ageing studies, the temporal evolution of neuropsychological tests is used to assess the cognitive abilities trend among the older adults \cite{alperovitch_les_2002}.

Marker data repeatedly measured over time in longitudinal studies have several characteristics that raise analytical challenges. First, the structure of the data induces a strong within-individual correlation which implies a variability across individuals. Second, the data collection process \cite{lokku_summarizing_2020}, often heavy in medical studies, usually leads to sparse measurements and individual-specific timings. For instance, in epidemiological cohorts of chronic diseases, it is not rare to have data every 2 to 5 years only \cite{boscardin_longitudinal_1998, ilic_prostate_2018, alperovitch_les_2002}. Third, participants may leave the study before the end because they died or because they refused to undergo the next visit which leads to dropout \cite{hogan_handling_2004}. These early truncations of the longitudinal process are often linked to the health status of the participant, and thus are likely informative. Fourth, marker data collected in health studies are systematically noisy observations of the true underlying process of actual interest. Finally, the marker trajectories may exhibit different shapes, requiring flexible modeling tools of time trends. Figure \ref{fig:applimiss} illustrates the characteristics of such marker data in the context of four neuropsychological tests collected in a cognitive aging study with the error-prone trajectories and sparse measurements (left panels), the individual-specific timings (middle panels) and the missingness patterns (right panels).

Mixed-effect models (MM) are the standard method to analyze longitudinal data in health studies \cite{verbeke_linear_1997}. They extend the classic linear regression models to correlated data by modeling both marginal trajectories using fixed effects and the individual-specific deviations using individual regression coefficients called the random effects. MM split the observations into the observation error and the underlying data generation process defined in continuous time, thus handling any measurement time and accounting for serial correlation. Any basis of time functions can be considered, including splines or polynomials, for flexible modeling. As estimated in the Maximum Likelihood framework, MM estimates are robust to missing data as long as the missingness mechanism is predictable by the observed data (mechanism also known as Missing at Random) \cite{rubin_inference_1976,little_statistical_1987}. MM thus handle most challenges faced with the longitudinal data collected in health data. However, MM are parametric tools; the distribution of both the error and the random effects must be specified by the user. Also, in some specific cases, the flexibility put into the MM may come with a great computational cost, particularly when the model is nonlinear and includes many random effects because of an heavy numerical integration in the log-likelihood calculation \cite{davidian_nonlinear_2003}.

Repeated measures of markers are observations of a continuous-time process and as such they could also be considered as sparse and irregular functional data. Indeed, classic functional data is a collection of noisy observations from an unknown random function $f$ on a regular and temporally dense grid of time. In functional data analysis, one of the main goal is to describe and summarize the temporal evolution of these trajectories. One of the most well-known method to do so is the functional principal component analysis (FPCA) \cite{ramsay_functional_2005, wang_review_2015}. Similarly to classic principal component analysis, it projects functional data into a lower dimensional functional space defined by a functional basis. In this functional space, each participant has its own set of coordinates called functional principal component scores. In health studies, FPCA is also very appealing because it allows, with a fully non parametric approach, to describe and summarize temporal trajectories in a parsimonious way using only a few principal functional components and individual contributions to these components. However, to be suitable for the analysis of longitudinal data stemmed from health studies, functional analysis methods should necessarily handle the challenges of sparsity, irregularity of the observation grid and missing data. Previous work have proposed FPCA for sparse longitudinal data \cite{yao_functional_2005}. Furthermore, FPCA should work in theory with irregular observation grids although current implementations in \texttt{R} packages are not always flexible enough. The main unsolved challenge is the behaviour of FPCA techniques with missingness mechanisms. No theoretical or numerical evaluation has been sought. While FCPA is expected to behave well in case of completely random missingness since it constitutes a specific type of irregular grid, the behaviour of the FPCA to other types of missingness still deserve to be investigated, especially when working with longitudinal data from health studies where informative missingness is common. The aim of this paper is to evaluate the robustness of FPCA to various schemes of missing data in an empirical simulation study and to compare its performances to the classic mixed-effect model approach.

Section \ref{sec::3Cstudy} briefly introduces the motivating data example with repeated evaluations of cognitive functioning in a population-based aging cohort. In Section \ref{sec::methods}, the statistical framework - in particular the missing data mechanisms - is described and both the mixed-effect models and the FPCA are introduced. Results from two empirical simulation studies are reported in Section \ref{sec::simus}. The two methods are then applied in Section \ref{sec::appli} to describe the trajectories of four cognitive functions in years preceding clinical dementia onset in contrast with the trajectories of matched controls. Our findings are summarized and discussed in Section \ref{sec::discussion}.

\section{The Three-City Study}
\label{sec::3Cstudy}

The Three-City Study (3C Study) is a French observational cohort study started in 1999 aiming at exploring the link between vascular diseases and dementia in the older adults \cite{alperovitch_les_2002}. Subjects aged $65$ years and older at baseline were recruited in three French cities and followed-up every 2 to 3 years during 17 years. At each visit, the cognitive functioning was measured by a battery of cognitive tests, and a diagnosis of dementia was established according to a three steps procedure: a screening based on neuropsychological performance; a clinical evaluation performed by a neurologist; and a final diagnosis made by an independent committee of neurologists (see \cite{alperovitch_les_2002} for details).

Our motivating aim was to describe the cognitive trajectories of participants diagnosed with dementia and contrast them with those expected during natural aging. To this end, we built a nested case-control study from the 3C Study. We included the $174$ incident dementia cases diagnosed in the Bordeaux Center with at least one measure at each cognitive test, and individually matched them in density of incidence with $156$ controls free of dementia at the time of diagnosis of the case. The matching was done according to sex, presence of at least one APOE-$\epsilon$4 allele (main genetic risk factor of dementia), level of education, age at diagnosis with a $3$-year margin, and length of follow-up at diagnosis (in years). We removed the cognitive measures occurring 5 years after the diagnosis (considered as time 0) as most cases were not able anymore to undergo a large cognitive battery. 

We focused on four psychometric scores that evaluate different cognitive aspects: the Mini-Mental State Examination score (MMSE) \cite{folstein_mini-mental_1975} which evaluates global cognitive functioning (with range 0 to 30), the Benton Visual Retention Test score (BVRT) \cite{benton_revised_1963} which measures visual memory (with range 0 to 15), the 30-second Isaacs Set Test (IST) score \cite{isaacs_set_1973} which assesses verbal fluency (with range 0 to 73) and the number of correct moves per minute at the Trail Making Test score part A (TMTA) \cite{reitan_validity_1958} which assesses executive functioning and attention. Note that the TMTA was not assessed at the second visit.

\begin{figure}
    \centering
    \includegraphics[width=\linewidth]{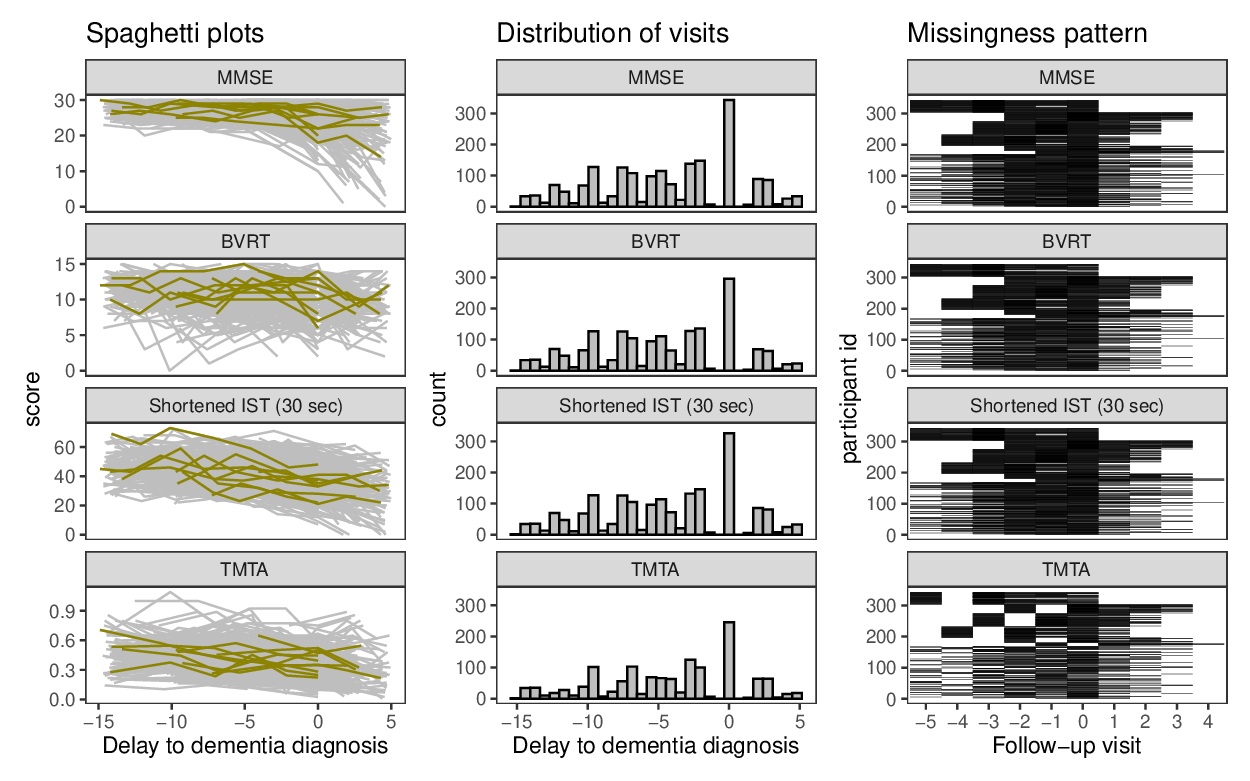}
    \caption{Description of the repeated data in $174$ dementia cases and $174$ matched controls (N=$348$) from the 3C study according to the delay (in years) to dementia diagnosis for the cases, and the delay to the dementia diagnosis of the matched case for controls (with $0$ corresponding to the dementia diagnosis visit of the case). Are reported the individual observed trajectories over time to diagnosis of four cognitive scores (MMSE, BVRT, IST30 and TMT-A) with $10$ randomly selected individual trajectories emphasized in green (first column); the distribution of the times of measurement (second column) and the missingness pattern vizualisation for each individual at each visit (third column). A black segment means the score is observed, a white segment means no observation is available. The strong structure on the missingness pattern is a consequence of the matched nature of the data.}
    \label{fig:applimiss}
\end{figure}

Figure \ref{fig:applimiss} describes the sample with the individual cognitive trajectories, the timings and patterns of missingness, thus illustrating the sparsity, irregularity in timings and missing data usually encountered in cohort studies.

\section{Methods}
\label{sec::methods}

\subsection{Sparse and irregular repeated measures data}

We assume that we observe repeated measures of a longitudinal marker over time $t$ for each unit $i$ of a sample of size $N$. These observations $y_{ij}$ also noted $y_i(t_{ij})$ with $j=1,\dots,n_i$ lead to an individual vector of observations $\mathbf{y_i} = (y_{i1}, \dots, y_{in_i})^{\top}$ where $n_i$ denotes the number of observations for unit $i$. The observations $y_{ij}$ are measurements with error of the underlying process of interest $y_i^{\star}(t)$ so that $y_{ij} = y^{\star}(t_{ij}) + \varepsilon_{ij}$ where $\varepsilon_{ij}$ denotes the independent random observation error.

Classically, in the functional data framework, the observation grid $\mathbf{t}=(t_1,\dots,t_{n})^{\top}$ is dense, regular over time (e.g. hourly measures, daily measures) and the same across the $n$ units. In contrast, in health studies on humans (cohorts, clinical trials), the data collection is frequently done with a lower frequency and the visit times $\mathbf{t_i}=(t_{i1},\dots,t_{in_i})^{\top}$ may vary across individuals. For example, in the 3C cohort data, there may be variations of several months or even years across individuals around the theoretical visits every 2 to 3 years because collecting cognitive test results implies an heavy and expensive observation process. Additionally, it is very common to have missing data which leads to different size of observation grid. Such a design defines sparse and irregular functional data. 

\subsection{Missing data mechanisms}

We use the classical formalism of missing data for a marker of interest as introduced by Little and Rubin (1987) \cite{little_statistical_1987}. We denote $Y$ the set of all marker data which includes those that are observed $Y^o$ and those that are missing $Y^m$. The binary variable $R$ is the indicator variable of observation with $R_{ij}=1$ if $Y_{ij}$ is observed, and $0$ otherwise. Little and Rubin (1987) \cite{little_statistical_1987} distinguishes three mechanisms of missingness. Data is called Missing Completely At Random (MCAR) when the probability that the marker is missing does not depend on the marker at all,
\begin{equation*}
P(R|Y^o,Y^m) =  P(R).
\end{equation*}
It happens for instance when a cognitive test is unavailable at a certain visit because this test was not scheduled in the protocol for this visit (as this is the case for the TMTA at the second visit in 3C). Data is called Missing At Random (MAR) when the probability that the marker is missing depends on the observed values of the marker,
\begin{equation*}
P(R|Y^o,Y^m) =  P(R|Y^o).
\end{equation*}
It happens for instance when a participant refusal to pass a cognitive test depends upon his/her results at previous visits. Finally, data is called Missing Not At Random (MNAR) when it can also depend on the unobserved values of the marker,
\begin{equation*}
P(R|Y^o,Y^m) =  P(R|Y^o,Y^m).
\end{equation*}
This can happen for instance when a participant refuses to pass the test because he/she feels that his/her cognitive level has dropped since last visit.

Missing data can be divided into two main types: intermittent missing data, when a measurement is missing during the follow-up but other measurement can be collected later on, and dropout data which happens when any visit after the dropout visit will be missing. In this work, we are interested in both types of missing data.

\subsection{Mixed-effect models}

Mixed-effect models have been introduced by Laird and Ware (1982) \cite{laird_random-effects_1982} and since then have been widely used to model repeated measures collected in longitudinal health studies. We focus on a mixed model that assumes a continuous marker with zero-mean independent Gaussian errors $\varepsilon_{ij}$ so that $y_{ij} = y_i^{\star}(t_{ij}) + \varepsilon_{ij}$. The underlying value of the marker at time $t$ is then modelled as follows:
\begin{equation}
	\label{eqn::lmm}
	y_i^{\star}(t) = g(t, X_i(t), \bm{\beta}, \bm{b_i})
\end{equation}
where $X_i(t)$ denotes a set of time-dependent covariates, $\bm{\beta}$ denotes the vector of fixed parameters, 
$\bm{b_i}$ denotes the vector of random effects with $\bm{b_i}\sim\mathcal{N}(0,\bm{B})$ and $g$ a basis of functions that describes the shape of the marker trajectory. The observation error $\varepsilon_{ij}\sim\mathcal{N}(0,\sigma^2)$ is assumed to be independent from the vector of random effects. In the remainder of this work, we concentrate on linear mixed-effect models where $g(t, X_i(t), \bm{\beta}, \bm{b_i}) = \sum_{k=1}^p (\beta_k+ b_{ik})g_k(t, X_i(t))$ with $(g_1,\dots,g_p)$ a $p$-basis of a priori specified functions of time, most often splines or polynomials.

Mixed-effect models have been extended to handle different types of data: binary or count data using generalized linear mixed model theory, or continuous data with strong ceiling or floor effect which is very common among neuropsychological scores \cite{proust-lima_sensitivity_2007, proust-lima_misuse_2011}. For the latter, a parametric monotonic function is applied to $Y$ in Equation \ref{eqn::lmm} to normalize the marker data.

The mixed-effect model is generally estimated by maximizing the log-likelihood: 
$$\ell(\bm{\theta}) = \sum_{i=1}^N \log \int f(Y_i|\bm{b_i})f(\bm{b_i})d\bm{b_i}.$$
where $\bm{\theta}$ denotes the model parameters $(\theta=(\bm{\beta}, \text{vec}(\bm{B}),\sigma)^\top)$, $f(\bm{Y_i}|\bm{b_i})$ is the density function of the outcome conditional to the random effects and $f(\bm{b_i})$ is the density of the random effects. When the model is linear in the random effects, the integral has an analytical solution. In other cases, the log-likelihood computation often implies to numerically approximate the integral \cite{pan_gauss-hermite_2003}.

Fitted values of existing observations or predictions of new data are easily derived from a fitted mixed model using the a posteriori distribution of the random effects given the individual observations. In linear mixed-effect models, it is directly computed using conditional expectation properties of multivariate normal variables while for nonlinear models, it is approximated by the mode of the posterior distribution \cite{laird_random-effects_1982, molenberghs_linear_2000}.

Despite the possible use of flexible families of time-functions (e.g., splines), a MM remains a parametric model with a distribution for the random effects and for the random error to be chosen (often Gaussian). Additionally, some assumptions on the variance structure of the random effect might be needed when the number of random effects increases to avoid a too large number of covariance parameters. 

One essential feature of the MM is its robustness to MAR data, which ensures an unbiased estimator under MAR \cite{rubin_inference_1976}. When the missing data are suspected to be MNAR, that is when the missing mechanism also depends on unobserved data, a joint model of the missingness mechanism is required. This can be done with a logistic model for intermittent missing data or a survival model for dropout \cite{diggle_informative_1994, rizopoulos_joint_2012}. However, in any case, the model requires a correct specification of the missing data mechanism to yield robust estimates \cite{thomadakis_longitudinal_2019}. For instance, if the instantaneous risk of dropout is assumed to be associated with the marker level, the following proportional hazard model can be jointly estimated with the mixed model in (\ref{eqn::lmm}):
\begin{equation*}
\alpha_i(t) = \alpha_0(t)\exp(\tilde{X}_{ij}^{\top}\bm{\gamma} + y_i^{\star}(t)^{\top}\bm{\eta})
\end{equation*}
with $\alpha_0(t)$ the baseline risk function, $\tilde{X}_{ij}^{\top}$ a set of covariates, $\bm{\gamma}$ and $\bm{\eta}$ two vectors of parameters to be estimated.

\subsection{Functional Principal Component Analysis}

Functional Principal Component Analysis has been introduced by Besse and Ramsay (1986) \cite{besse_principal_1986} and since then has been used for describing, modeling, predicting or classifying abundant temporal series \cite{shang_survey_2014, wang_review_2015}. As for mixed-effect models, we consider additive random errors such that observations $y_{ij} = y^{\star}_i(t_{ij}) + \varepsilon_{ij}$ with $y^{\star}_i(t_{ij})$ the underlying process of interest. 
In functional data, the vector $\bm{y^{\star}_i} =\{y^{\star}(t_{ij})\}_{j=1,...,n_i}$ for unit $i$ is considered as a random realization of an unknown underlying function. Consequently, the whole set of longitudinal trajectories $\bm{y^{\star}}=(\bm{y^{\star}_1}, \dots, \bm{y^{\star}_N})^{\top}$ is a collection of realizations of an unknown random function $f$ assumed to be smooth. This unknown function has mean $\mu(t)$ and covariance function $G(s,t)=\text{cov}(f(t),f(s))$. FPCA is essentially based on the Karhunen-Loève decomposition \cite{karhunen_zur_1946, loeve_fonctions_1946} which states that the underlying unknown smooth function $f$ can be decomposed into:
\begin{equation}
f_i(t) = \mu(t) + \sum_{k=1}^{\infty} \xi_{ik}\phi_k(t)
\label{eqn::fpca1}
\end{equation}
where the index $i$ in $f_i$ emphasizes the fact that the random realisation of $f$ is subject specific, $\xi_{ik}$ and $\phi_k$ are the eigenvalues (or functional principal component scores) and eigenfunctions from the decomposition of the covariance operator $G$, respectively. From this decomposition, dimension reduction is straightforward by limiting the sum in (\ref{eqn::fpca1}) to the first $K$ components only:
\begin{equation}
f_i(t) = \mu(t) + \sum_{k=1}^{K} \xi_{ik}\phi_k(t).
\label{eqn::fpca2}
\end{equation}

From the noisy data $y_{ij}$, local linear smoothers \cite{ramsay_functional_2005, yao_functional_2005} are used to estimate the mean function $\mu$ and the covariance function $G$. The estimated covariance surface is discretized to make spectral decomposition of the covariance operator easier and from that, eigenfunctions are obtained \cite{rice_estimating_1991, wang_review_2015}. In the case of dense functional data, the scores can be estimated using a numerical integration \cite{ramsay_functional_2005} which is not feasible in the case of sparse data. Functional Principal Components Analysis through Conditional Expectation (PACE) \cite{yao_functional_2005} has been proposed as an alternative way of computing the eigenvalues in case of sparse data. Using properties of conditional Gaussian multivariate density, an expression for $\mathbb{E}(\bm{\xi_i}|\bm{Y_i})$ is obtained, very close in spirit to what is done in linear mixed-effect models. From that, by plugging in the estimated eigenvalues, eigenfunctions, covariance and mean function, an estimate of the score can be computed. In PACE, the scores are assumed to be Gaussian but the method is robust to violation of this assumption \cite{yao_functional_2005}. 

When using FPCA, the determination of $K$ in \eqref{eqn::fpca2} is critical. Apart from rare exceptions, $K$ is determined from the data by optimizing a statistical criterion such as percentage of variance explained, AIC, or BIC where cross-validation can be used. Fitted values of  existing  observations or predictions of new data can be easily computed by plugging into (\ref{eqn::fpca2}) the estimated $\hat{\mu}(t_{ij})$ and $\hat{\phi}_k(t_{ij})$, and the individual eigenvalues $\hat{\xi}_{ik}$ computed using the PACE algorithm.

Many \texttt{R} packages have implemented FPCA for functional data but only a few are suited to the case of sparse and irregular functional data, in particular during the prediction step for new units. For example, \texttt{fda} \cite{fda} does not implement PACE and requires the same observation grid for all units which is not realistic in many health contexts. Package \texttt{MFPCA} \cite{MFPCA} is suited to sparse and irregular functional data but it struggles to make predictions on new units with different observation grids. We identified two packages for the estimation of FPCA in case of sparse and irregular data: \texttt{face} \cite{face} and  \texttt{fdapace} \cite{fdapace}. Their major difference lies in the specification of the smoothing used to estimate the functional components: \texttt{fdapace} uses local weighted polynomial smoothing while \texttt{face} uses P-splines smoothing. We thus considered the two implementations in our work.

Contrarily to linear mixed-effect models, no theoretical justification exists regarding FPCA robustness to missing at random data, and no empirical study has explored this issue so far.

\section{Simulation study}
\label{sec::simus}

We numerically evaluated the performances of FPCA to analyze sparse, noisy and irregular repeated data in the presence of dropout with (i) a comparison of the predicted values with those obtained by mixed models considered as the gold standard approach, and (ii) an evaluation of the robustness of FCPA estimations to dropout. We followed the recommendations of Morris et al. (2019)\cite{morris_using_2019} to plan and report these two simulation studies.

\subsection{Comparison between FPCA and MM}
\label{subsec::FPCAvsLM}
\subsubsection{Aims} 

The first simulation aimed to contrast the performances of MM and FPCA to predict missing values under different scenarios of dropout with different implementations of MM and FPCA.

\subsubsection{Data generation mechanism}

We generated individual repeated data in samples of $N=700$ participants. The sparse visit process was generated from a fixed grid of visits between $t=0$ and $t=12$ with three levels of sparsity in the visits times: a measure every $3$, $2$ or $1$ years leading to a maximum of $5$, $7$ or $13$ observations per individual.

To mimic a more realistic observation window, the individual visit times $t_{ij}$ were obtained by adding a random uniform noise around the theoretical visits for each observation ($\pm~1.5$, $1$, $0.5$ years for visits every $3$, $2$ and $1$ years, respectively). This uniform noise adds a first level of irregularity to the observation grid by making it specific to each individual. The repeated marker data at each visit time $t_{ij}$ were generated according to a mixed model with a nonlinear shape over time captured by a five-parameter logistic model as implemented in the \texttt{nlraa} package \cite{nlraa}:
$$y^{\star}_i(t_{ij}) = y_{\infty i} + \frac{y_{0i} - y_{\infty i}}{(1 + \exp(\alpha_i \times (\log t_{ij} - \log \tau_{i})))^{\theta_i}}$$
where $y_{0i} = 10 + b_{i1}$ and $y_{\infty i} = 35 + b_{i2}$ are the individual-specific asymptotic values of $y$ for 0 and infinite times, respectively; $\tau_{i} = 10 + b_{i3}$ is the time when the mean value of $y$ is attained (only if $\theta_i = 1$), i.e. $y(\tau_i) = (y_{0i} + y_{\infty i})/2$, $\alpha_i = 5 + b_{i4}$ controls the steepness of the transition and $\theta_i = 1 + b_{i4}/3$ is an asymmetry parameter. Individual-specific trajectories were generated assuming random-effects $b_i=(b_{1i}, b_{2i}, b_{3i}, b_{4i})^{\top}\sim\mathcal{N}(0,B)$ with $B$ a diagonal $4\times 4$-matrix with diagonal $(3,5,2,2)^{\top}$. 
The prone-to-error observations were derived by adding an independent homoscedastic Gaussian error $\varepsilon_{ij} \sim\mathcal{N}(0,3^2)$ such that: $y(t_{ij})=y^{\star}_i(t_{ij}) + \varepsilon_{ij}$. 

At this stage, we have a complete set of repeated measures for each participant up to the administrative censoring at year 12. We added a second level of irregularity to the observation grid by truncating the observation process at an individual time of dropout (i.e., early leave from the study), making the observation grid length subject-specific. We considered a dropout with two intensities ($30\%$ and $60\%$) and the six following missing data scenarios (see supplementary material for details):
\begin{itemize}
    \item MCAR: the participant drops out at $t_{ij}$ with a probability determined by a logistic model with $t_{ij}$ as predictor;
    \item fixed MAR: the participant systematically drops out at $t_{ij}$ if $Y_{i(j-1)}$ is above a threshold $\nu$;
    \item threshold MAR: the participant drops out at $t_{ij}$ with a probability determined by a logistic model with the indicator that $Y_{ij-1}>\nu$ as predictor;
    \item increasing MAR: the participant drops out at $t_{ij}$ with a probability determined by a logistic model with $Y_{ij-1}$ as predictor;
    \item threshold MNAR: the participant drops out at $t_{ij}$ with a probability determined by a logistic model with the indicator that $Y_{ij}>\nu$ as predictor;
    \item increasing MNAR: the participant drops out at $t_{ij}$ with a probability determined by a logistic model with $Y_{ij}$ as predictor;
\end{itemize}

\begin{figure}
    \centering
    \includegraphics[width=\linewidth]{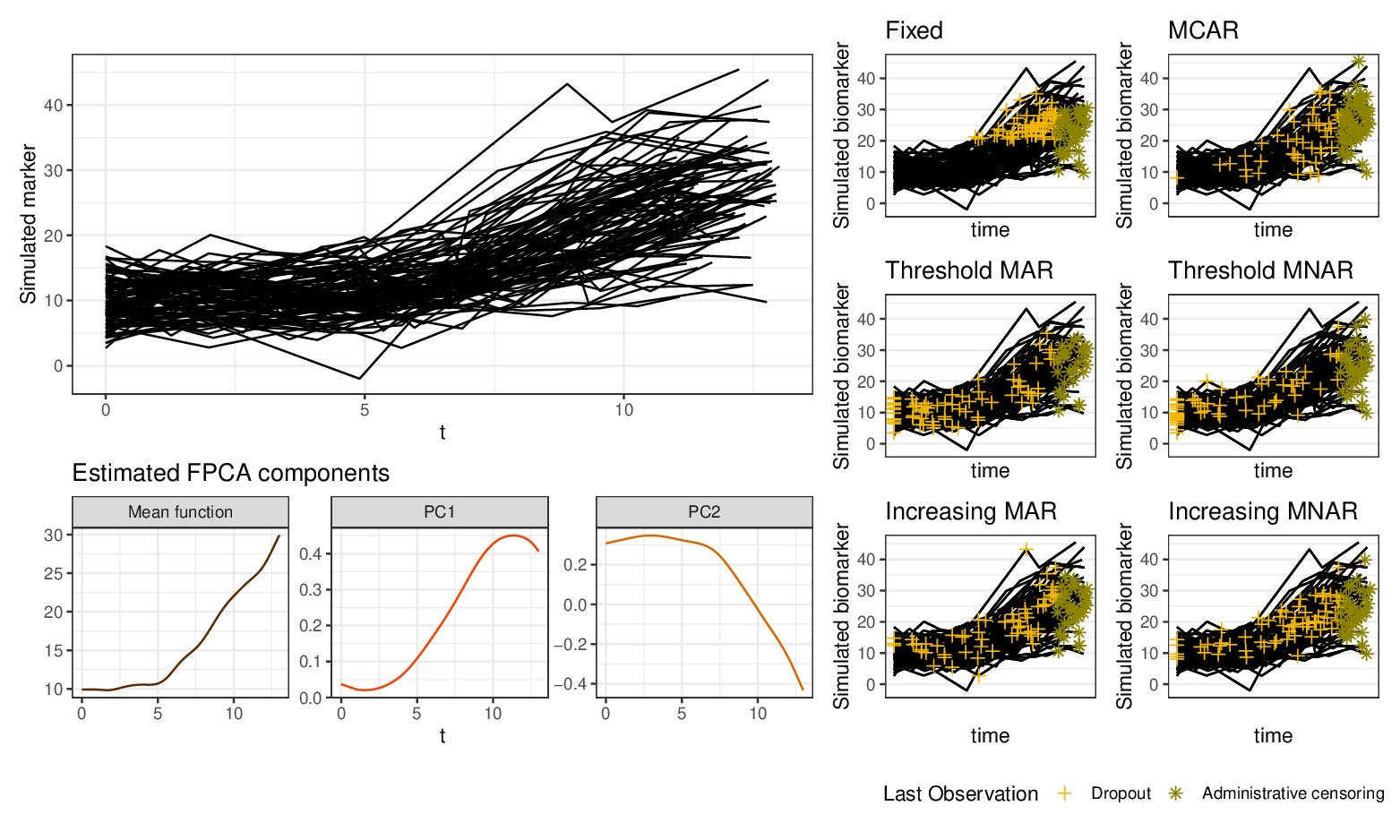}
    \caption{Top-left and right panels: generated individual longitudinal data according to a mixed model for $n=100$ randomly selected participants prior to dropout (top-left) and after 6 different generated mechanisms of dropout with rate $0.6$ (right). Bottom-left: the mean function and the two functional principal components that can be derived from the top-left individual trajectories using a FPCA.}
    \label{fig:genedata}
\end{figure}

The data generated as described above satisfies the three main challenging characteristics of longitudinal data usually encountered in epidemiological studies: sparsity with three levels of visit frequency, error-proneness with a Gaussian noise on the repeated measures of the marker, and irregular time grid both in visit occurrence and grid length. The generated longitudinal data of 100 randomly selected individuals are displayed in Figure \ref{fig:genedata}. The right panels illustrate the two types of early truncation of the longitudinal process: the dropout under the above mechanisms (with a rate of $0.6$) and the administrative censoring at year 12.
 
\subsubsection{Estimands}

Each sample was separated into a training set of $200$ participants and a test set of $500$ participants. The estimands of interest were the predictions of the marker values $y_{ij}$ in the test set whether they were observed or missing (true generated values were known from the generation procedure). These predictions were computed from the estimations of FPCA and MM on the training set.

\subsubsection{Methods}
\label{subsec::methods}
For each simulation setting, that is for a fixed frequency of visits (every $3$, $2$ and $1$ years), for a fixed rate of dropout ($30\%$ or $60\%$), for a fixed missing data scenario (MCAR, fixed MAR, threshold MAR, increasing MAR, threshold MNAR, increasing MNAR), we estimated the following models:

\begin{itemize}
    \item Linear mixed-effect models (including a random effect on the intercept and on each time function) with time trend: modelled as a quadratic trend (LMM\_quad);  modelled as a cubic trend (LMM\_cub); approximated by natural splines with 2 internal knots at the terciles (LMM\_spl\_quant); approximated by natural splines with 2 equidistant internal knots (LMM\_spl\_equi). 
    \item Functional Principal Component Analysis with the number of principal components chosen such that the percentage of explained variance reached $90\%$ (FPCA\_fve90\_fdapace) or $99\%$ (FPCA\_fve99\_fdapace) in \texttt{fdapace} or $99\%$ (FPCA\_face\_fve99) in \texttt{face}.
    \item Shared random effect joint model (JM) in which the longitudinal submodel followed a quadratic time trend (random intercept and random slope on each time function) and the survival submodel was a proportional hazard model with a baseline risk function approximated by B-splines (3 internal knots at the quartiles) and a dependence through the current underlying value of the marker. Note that the joint model was only estimated in the two MNAR scenarios (threshold MNAR and increasing MNAR) as a \emph{gold standard} comparator for the MNAR situation.
\end{itemize}

The linear mixed-effect models were estimated using the \texttt{lcmm} package \cite{lcmm} (function \texttt{hlme}) and predictions were obtained using the \texttt{predictRE} function. The joint models were estimated using the \texttt{JM} package \cite{JM}, and predictions computed with the \texttt{predict} function. FPCA models were estimated with the \texttt{fdapace} package \cite{fdapace} and the \texttt{face} package \cite{face}. In contrast with previous methods, prediction was only possible for times available in the training set. We thus discretized the time window (with a step of $0.5$) to ensure that every discretized $t_{ij}$ in the test set also appeared in the training set, and used the internal \texttt{predict} function of \texttt{fdapace} and \texttt{face}. 

\subsubsection{Performance measure}

For each setting, the procedure was replicated on 1000 replicates. The performances were evaluated with the root mean square error (RMSE) by separating: (i) the RMSE computed on the actual observations of the test set (i.e., non-missing due to dropout), and (ii) the RMSE computed on the missing data of the test set (after time of dropout). 

To make all the RMSEs comparable across models within a fixed scenario, we considered standardized RMSEs. To do so, we first estimated a reference flexible model in the complete training set (i.e. without any missing data). This reference model was chosen as a mixed model with B-splines time trend with $3$ internal knots placed at the quartiles. We computed the RMSE of this model on test data (either missing data only, or observed data only) and defined it as the reference RMSE. Then, for every other model estimated on the training set with missing data, we computed their RMSEs on test data (either missing data only, or observed data only) and standardized them by dividing them by the corresponding reference RMSE. 

\subsubsection{Results}

The standardized RMSEs of each model computed on the missing test data are reported in Figure \ref{fig:simusRMSE} (see Supplementary Figures 2, 3 and 4 for the observed test data) for four missing data scenarios (MCAR, fixed MAR, increasing MAR and increasing MNAR), two rates of dropout ($30\%$ and $60\%$), and the three frequencies of sparse visits. 

\begin{figure}
    \centering
    \includegraphics[width=\linewidth]{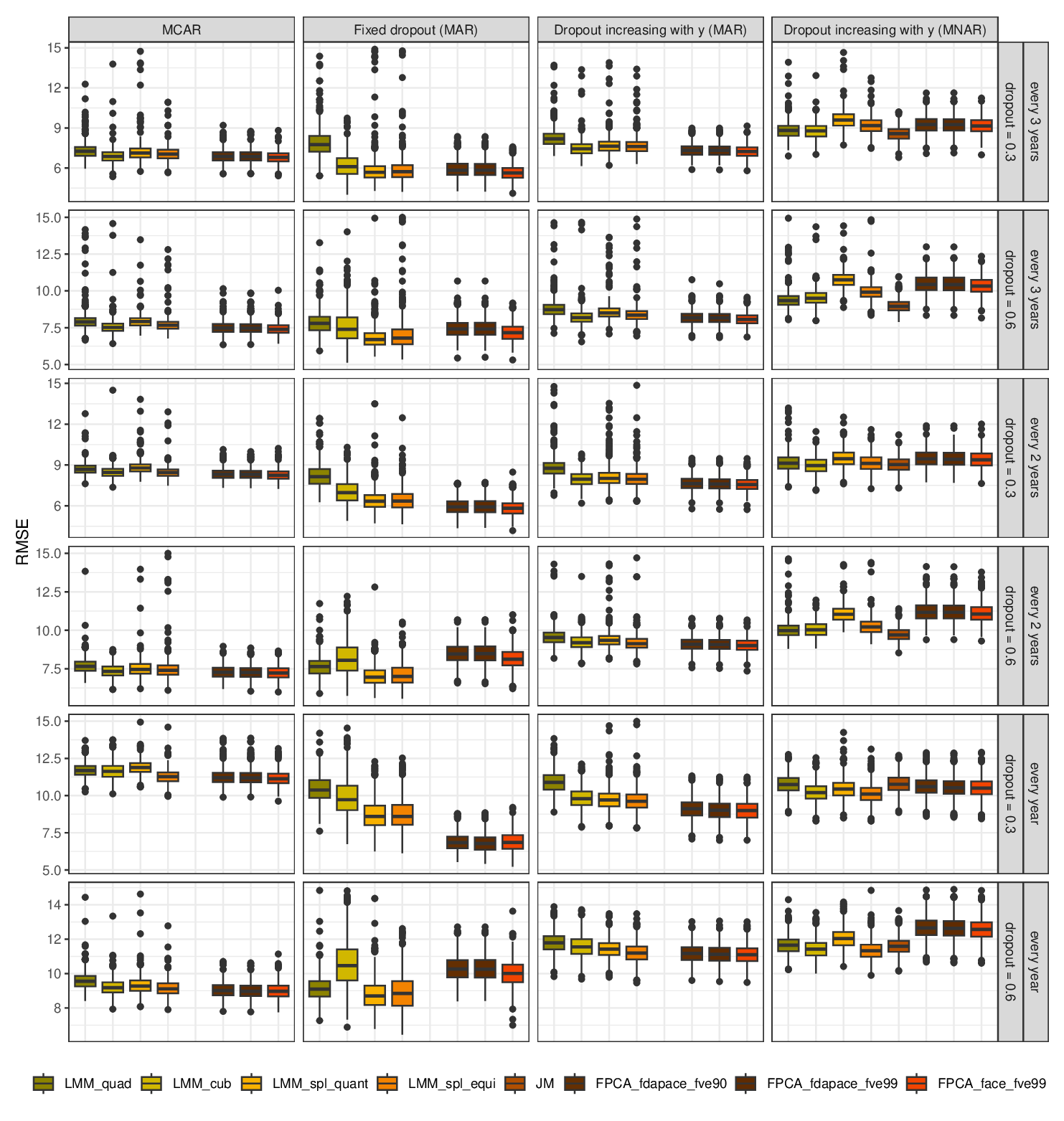}
    \caption{Standardized RMSEs computed on missing test data over $1000$ replicates for four missing data scenarios (MCAR, fixed MAR, increasing MAR and increasing MNAR from left to right), two rates of dropout ($30\%$ and $60\%$) and three frequencies of sparse visits (every 3, 2 and 1 year from top to bottom). Each boxplot summarises the RMSEs of a model relative to the one of a reference model estimated on complete data (from left to right): LMM\_quad, LMM\_cub, LMM\_spl\_quant, LMM\_spl\_equi, JM (only for the MNAR case), FPCA\_fdapace\_fve90, FPCA\_fdapace\_fve99, FPCA\_face\_fve99.}
    \label{fig:simusRMSE}
\end{figure}

An overall look at mixed models RMSEs shows that flexible mixed models, those based on spline parametrization, rank systematically among the best for every missing data scenario, dropout rate and observation grid. As expected, virtually the same standardized RMSE were obtained for MM and FPCA in the MCAR case, whatever the frequency of visit and the rate of dropout. With a fixed MAR dropout after an observation above a given threshold, the MM relying on splines showed the lowest standardized RMSEs. With a 30\% rate of dropout, those of FPCA were comparable or even lower when data were denser (every year). However, with a 60\% rate of dropout, the performances of FPCA dropped, probably due to the sudden and local loss of information. With a MAR dropout that increases as the last observed marker value increased, the dropout is more spread along the observation window and FPCA lead to very competing standardized RMSEs compared to MM with splines time trends. This is true whatever the rate of dropout, and even provide slightly better fit for denser data and lower rate of dropout, underlying their flexibility to capture complex trends. With a MNAR dropout that increases as the simultaneous marker level increases, MM and FPCA models show larger standardized RMSEs compared to the more appropriate JM. However, again FPCA shows worse behaviour than MM when the rate of dropout is high (60\%). The two implementations of the FPCA gave overall the same results. We observed an expected but very tiny improvement when targeting $99\%$ of explained variance rather than only $90\%$. The different specifications of the components between \texttt{face} and \texttt{fdapace} gave overall the same results in all scenarios. The fit of the spline-based FPCA (in \texttt{face}) was slightly better than the one of the local-polynomial based FPCA (in \texttt{fdapace}) when the dropout was systematic after the outcome reached a certain level (fixed MAR) but the differences remained negligible in all the other scenarios.

In summary, this simulation study suggests that, with sparsely and irregularly measured data, FPCA behave similarly as the flexible linear mixed-effect models in MCAR, MAR and MNAR scenarios. In some specific situations when combining a higher rate of dropout and very specific missing scheme (systematic dropout above a threshold), the FPCA seemed to perform slightly worse in capturing the entire underlying pattern. These results suggest that FPCA may constitute a suitable alternative to MM when the information loss is progressive, allowing the FPCA to retrieve the underlying trend.

The results for the threshold MAR and threshold MNAR cases, displayed in Supplementary Figure 1, lead to the same conclusions. 

\subsection{Robustness of FPCA to dropout}
\label{subsec::simusFPCA}
\subsubsection{Aims}
This second simulation study aimed to evaluate to which extent the components and scores of FPCA were correctly estimated in the presence of dropout.

\subsubsection{Data generation mechanism}

The data generation mechanism was exactly the same as in the first simulation study except that the samples were of size $N=200$ and the time function for the underlying marker trajectory was generated according to a two-component FPCA rather than a mixed model. 

The generating mean function and principal components, displayed in the bottom-left panel of Figure \ref{fig:genedata}, were chosen as those obtained after estimating a FPCA on a dataset simulated as in Section \ref{subsec::FPCAvsLM}. For each individual, we then randomly sampled Gaussian scores $\xi_i\sim\mathcal{N}(0,\Xi),$ (with $\Xi$ taken as the empirical variance of the scores derived from the FPCA) and applied equation \eqref{eqn::fpca2} to generate the functional data at each observed visit time. 

\subsubsection{Estimands}
The quantities of interest were the estimated mean function and the estimated functional principal components.

\subsubsection{Methods}
For each setting (rate of dropout, mechanism of dropout and frequency of visits), we applied a FPCA implemented in \texttt{fdapace} package \cite{fdapace} to the observed data, fixing the number of components to $2$.

\subsubsection{Performance measure}
We evaluated the relative bias between the estimated mean function or principal components, and their true counterparts on a fixed time grid over $1000$ replicates. Additionally, we visually compared the empirical distribution of the estimated mean function and functional principal components to the generated ones.

\subsubsection{Results}

\begin{figure}
    \centering
    \includegraphics[width=0.9\linewidth]{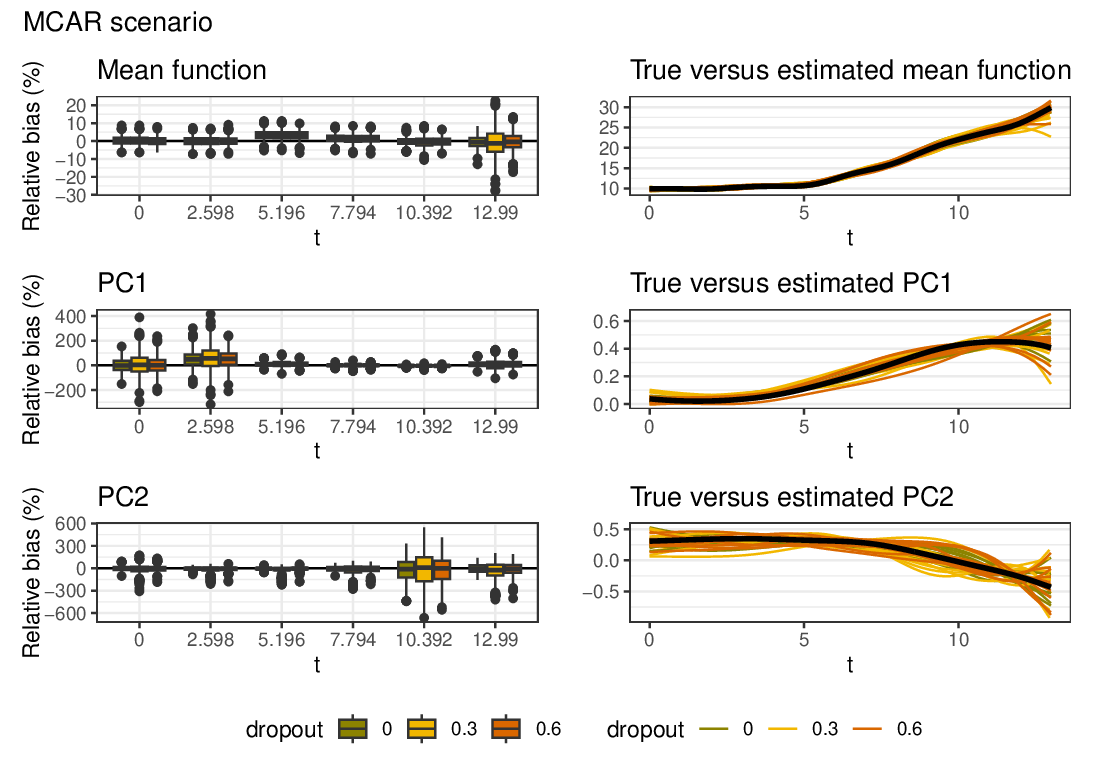}
    \includegraphics[width=0.9\linewidth]{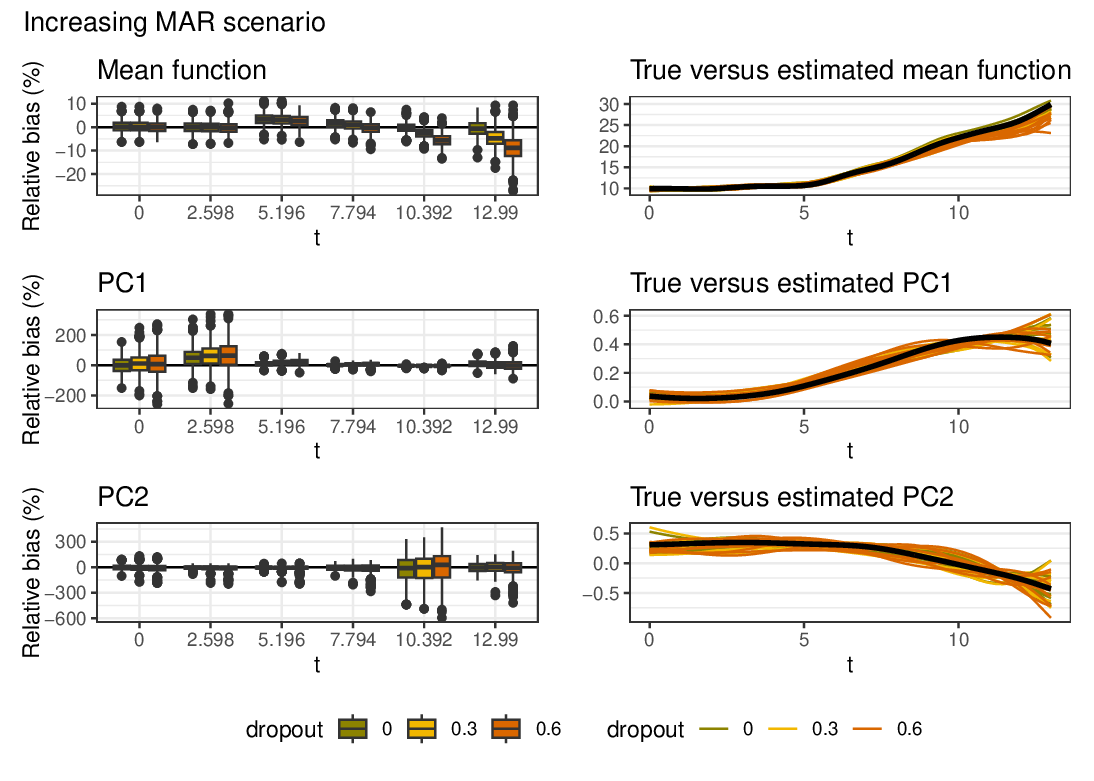}
    \caption{Relative bias over $1000$ replicates (left panel), and true plus $10$ randomly estimations (right panel) for the mean function and each of the two functional principal components in MCAR and MAR scenarios. The settings are a measure every 2 years for 12 years with a dropout rate of $0$ (no dropout), $30\%$ or $60\%$.}
    \label{fig:simusFPCA}
\end{figure}

The simulation results are reported in Figure \ref{fig:simusFPCA} for both the MCAR and increasing MAR scenarios when considering a measure every 2 years. As expected, in the MCAR scenarios (upper panels), there is no bias in the estimations whatever the dropout rate (0\%, 30\% or 60\%) and the estimated curves are close to the true generated mean and principal components. 
In the increasing MAR (lower panels) and the threshold MAR (Supplementary Figure 12) scenarios, the principal components are overall estimated without bias whatever the dropout rate (0\%, 30\% or 60\%) and the variability of the estimates does not substantially change with the dropout rate. The mean function in contrast shows a small bias at the end of the observation window that increases as the rate of dropout increases. Indeed the mechanism of dropout imposes a larger proportion of missing data by the end of the observation window. These observations are further illustrated by the $10$ randomly selected estimated curves which highlight the departure from the true mean function in the 30\% and 60\% dropout cases. With a fixed MAR dropout after an observation above a threshold, the FPCA performed poorly at the end of the observation window as expected due to the sudden and systematic loss of information. In the increasing and threshold MNAR scenarios (Supplementary Figures 13 and 14), the relative bias for the mean function is overall larger and especially at the end of the follow-up. The same results (Supplementary Figures 5-10 and 15-20) were observed for the other frequencies of observations.

\section{Application to cognitive trajectories in prodromal dementia}
\label{sec::appli}

We applied FPCA and compare it to MM to flexibly describe the trajectories of the four cognitive functions (global functioning, visual memory, verbal fluency, executive functioning) before clinical dementia, and contrast them with those of matched controls in the case-control sample nested within the 3C study. Each cognitive function was measured by a specific score: MMSE for global functioning, BVRT for visual memory, IST for verbal fluency and TMTA for executive functioning. The main characteristics of the nested case-control study sample are reported in Table \ref{tab::3C} and individual trajectories and missingness mechanisms are plotted in Figure \ref{fig:applimiss}.

\begin{table}
\centering
\begin{tabular}{lcc}
\hline
Variable & Cases & Controls \\
& (N=174) & (N=174)\\
\hline
Women & 129 ($74.1\%$) &  129 ($74.1\%$) \\
Index age$^*$  & $84.8$ $(81.6-88.9)$ & $84.8$ $(81.6-88.7)$ \\
Secondary school or higher & 106 ($60.9\%$) & 106 ($60.9\%$) \\
At least one APOE-$\epsilon$4 allele & 44 ($25.3\%$) & 44 ($25.3\%$) \\ 
Follow-up at diagnosis (in years) & $9.72$ $(6.78-12.2)$ & $9.70$ $(7.08 - 12.4)$ \\
MMS score at index age & $23$ $(21-25)$ & $28$ $(25-29)$ \\
\emph{\quad Number of individual measures of MMS} & $6$ $(4-7)$ & $6$ $(5-7)$ \\
BVRT score at index age & $9$ $(8-11)$ & $11$ $(9-12)$\\
\emph{\quad Number of individual measures of BVRT} & $5$ $(4-6)$ & $5$ $(4-7)$ \\
IST30 score at index age & $28$ $(22-34)$ & $39$ $(34-47)$ \\
\emph{\quad Number of individual  measures of IST30} & $5$ $(4-7)$ & $6$ $(4.25-7)$ \\
TMTA score at index age & $0.25$ $(0.21-0.35)$ &  $0.39$ $(0.31-0.48)$\\
\emph{\quad Number of individual  measures of TMTA} & $4$ $(3-5)$ & $4$ $(3-5)$ \\
\hline 
\end{tabular}\\
{\small{$^*$ Index age corresponds to the matching age at which cases were diagnosed with dementia, and controls were still free of dementia.}}\\
\caption{Summary statistics of demographic variables and cognitive scores in the case and control groups from the nested matched case-control (N=348, the 3C Study cohort). Are reported count (frequency) for categorical variables and median (interquartile range) for continuous data and number of measures.}
\label{tab::3C}
\end{table}

The analytical sample comprised $74\%$ of women in both groups; 61\% had at least finished secondary school and $25.3\%$ were carriers of at least one APOE-$\epsilon$4 allele. The median age at diagnosis was $84.8$ (interquartile range (IQR) = $81.6 - 88.7$) for cases. The matched index age for controls was $84.8$ (IQR = $81.6 - 88.8$). The four cognitive scores distribution implies a worst cognitive condition for cases with a median systematically lower and an interquartile range extending systematically below those of controls. The number of repeated measures per participant was similar between cases and controls and also among cognitive scores except for TMTA, which had fewer measures.

A separate analysis was conducted among cases and controls. For each subset and each cognitive score we estimated: a FPCA with a targeted percentage of explained variance of $99\%$, and a linear mixed-effect model in which the cognitive trajectory was modeled with natural cubic splines with two internal knots at the quartiles, and boundary knots at $-10$ and $2$ years to reduce the influence of observations beyond this range. With FPCA as implemented in \texttt{fdapace}, no control was possible for reducing the influence of time periods during which the information is sparser. For both methodologies, $95\%$ confidence intervals were computed: directly from the \texttt{lcmm} output based on the inverse of the negative Hessian matrix for the mixed models, and computing bootstrap confidence intervals with $500$ bootstrap samples with the function \texttt{GetMeanCI} for the FPCA.

\begin{figure}
    \centering
    \includegraphics[width=\linewidth]{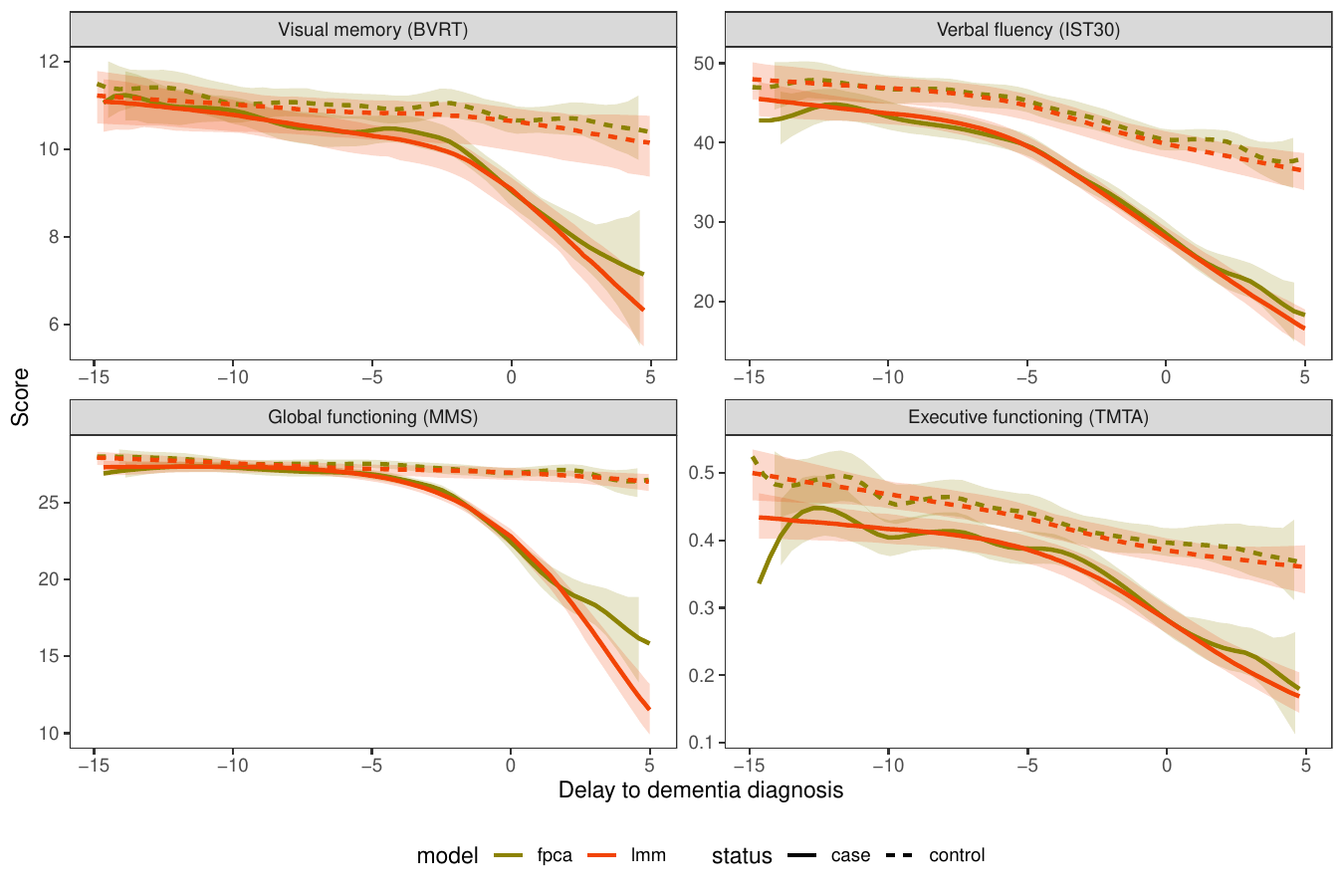}
    \caption{Marginal estimation from the linear mixed-effect model and mean function estimation from the FPCA and their $95\%$ confidence intervals for each cognitive marker among cases and matched controls of the nested case-control study from 3C cohort ($N=348$)}
    \label{fig:applires}
\end{figure}

The estimated marginal trajectories of the cognitive markers obtained by the two methods in each case/control group are displayed in Figure \ref{fig:applires}. Overall, the estimated mean trajectories and $95\%$ confidence intervals were very similar across methods except for slight differences observed at the end of the observation window. The predicted mean trajectories obtained from the linear mixed-effect model were however much smoother than those estimated by the FPCA. 

The mean trajectories of the four cognitive functions (visual memory, verbal fluency, global functioning and executive functioning) showed very different trends for controls and cases. Controls had a slow linear decline over time for all the cognitive functions. Far from dementia diagnosis, cases also had a slight linear decline, almost overlapped with the one of the controls for global functioning, very close for visual memory, and almost parallel but approximately $0.1$ points and $5$ points below for executive functioning and verbal fluency respectively. Then closer to diagnosis, each cognitive decline showed a sharp acceleration among cases: around $5$ years before diagnosis for verbal fluency, $3$ years  before diagnosis for executive functioning and $2.5$ years before diagnosis for visual memory and global functioning.

\section{Discussion}
\label{sec::discussion}

Initially developed for dense and regular functional data, FPCA has been recently extended to handle sparsity \cite{yao_functional_2005}. If FPCA can be used to impute missing data, the robustness of FPCA to the presence of missing data remained uncertain. In this manuscript, we have conducted an empirical simulation study to evaluate the robustness of FPCA in the presence of missing data caused by dropout. Our findings demonstrate that, in practical applications, FPCA is well-suited for analyzing longitudinal data, even when missing data follow a Missing at Random (MAR) mechanism, exhibiting comparable performance to linear mixed-effect models known for their robustness in such scenarios. This observation holds true across various MAR scenarios, observation grid densities, and dropout rates with exception for specific scenarios with the combination of very frequent and systematic dropout above a threshold. FPCA along with linear mixed-effect models do not seem to be robust to Missing Not at Random data. These results are based on two FPCA implementations: \texttt{fdapace} and \texttt{face}, but from our experience, \texttt{fdapace} is currently the easiest implementation to use with sparse and irregular functional data.

This empirical evaluation specifically targeted dropout, a particular category of missing data pattern wherein the absence of data at any point implies that all subsequent visits are also missing. The dropout scenario may induce a larger information loss than intermittent missing data where information from the longitudinal trajectory can be retrieved, even at the end of the window time. Hence, we anticipate the robustness to dropout found in this work to also apply to intermittent missing data. Note that it is important to distinguish missing data due to dropout and missing data truncated by death \cite{kurland_longitudinal_2009, rouanet_interpretation_2019}. In this work, we only focused on missing data due to dropout.

Our work was restricted to designs in which the observation times were non-informative. This is the case in the 3C study as in most cohort studies in which the visits are planned beforehand. When observation times depend upon the measured outcome (e.g., when participants consult each time their health condition worsens), specific functional data techniques have been very recently proposed \cite{weaver_functional_2023, xu_bias-correction_2024}, including the use of weights in the estimating equations of FPCA \cite{weaver_functional_2023}. We used the standard weighting scheme implemented in PACE algorithm in which each observation has equal weight \cite{yao_functional_2005, zhang_sparse_2016}. Alternatively, subject-specific weights could be applied for balancing the contribution of each subject\cite{weaver_functional_2023} in case of informative visit times. In the presence of informative dropout, no weighting technique has been developed so far in software to our knowledge. We leave for future works the study of weighted versions, including notably weighting schemes proposed in marginal models estimated by generalized estimating equations in presence of dropout or death \cite{rouanet_interpretation_2019}.

In our application study, without any assumption on the shape of the cognitive trajectories, both methods (based on MM and FPCA) showed very similar trends. Despite the absence of control on the FPCA degree of smoothing using \texttt{fdapace} \cite{fdapace}, the mean predicted trajectories were smooth over time except for the TMTA for which variations at the very beginning of the observation window suggested an overfitting (see Figure \ref{fig:applires}). Compared to other scores, TMT was not collected at the second visit of the cohort and fewer measures were available between -15 and -10 (see last row in Figure \ref{fig:applimiss}).

Cognitive trajectories exhibited a linear mean trend for controls throughout the window of interest. In contrast, the cognitive trajectories of cases, closely resembling those of controls distant from the diagnosis, exhibited a break with an accelerated decline as the diagnosis approached. This two-stage trend in pathological cognitive decline leading towards dementia had already been identified in cognitive ageing research \cite{jacqmin-gadda_random_2006, amieva_compensatory_2014}. This empirical study highlights the usefulness of FPCA to closely describe, summarize and predict longitudinal data. As a non-parametric approach, it avoids dependence upon strong assumptions and can adapt to any shape of trajectory. However, inference tools are not straightforwardly available with FPCA contrarily to MM which benefits from the maximum likelihood framework. For instance, parametric changepoint mixed models have been proposed to conduct inferential tasks to specifically explore this cognitive decline two-stage trend \cite{dominicus_random_2008, segalas_hypothesis_2019}.

A parallel between the mixed-effect model and the FPCA frameworks exists. The role of the individual random effects in the mixed-effect model is very similar to the role of the individual scores in the FPCA. They both represent individual deviations to the mean trajectory: the marginal model in mixed-effect model and the mean function in the FPCA. Moreover, under Gaussian assumption, the random effect prediction for linear mixed-effect model is based on Gaussian conditional expectation properties and its formula is very close to FPCA score's formula from the PACE algorithm. Despite this parallel, FPCA and more broadly any method from functional data analysis were initially developed for dense and regular functional data. This work joins others that explored how to handle missing data using multiple imputation techniques in functional regressions with functional data as the covariate\cite{rao_modern_2021} or as the outcome \cite{ciarleglio_elucidating_2022}. All these techniques open up possibilities regarding the utilization of functional data analytical tools in the context of sparsely and irregularly measured data increasingly available in health studies.

\section*{Acknowledgement}
This study received financial support from the French government in the framework of the University of Bordeaux's France 2030 program / RRI PHDS. Computer time for this study was provided by the computing facilities of the MCIA (\emph{Mésocentre de Calcul Intensif Aquitain}).

\section*{Additional information}
A supplementary material is available, providing details on the simulations scenarios and additional results. Code used for the simulation is shared on the Github repository \href{https://github.com/crsgls/fpcarobustness}{\texttt{fpcarobustness}}.

{\footnotesize{
\bibliographystyle{plain}
\bibliography{biblio.bib}}}

\end{document}